\documentclass[aps,prl,twocolumn,tightenlines,superscriptaddress,preprintnumbers,floatfix]{revtex4}
\usepackage{amsmath,graphicx}
\begin{document}
\title{Viscous Hydrodynamic Predictions for Nuclear Collisions at the LHC}
\author{Matthew Luzum}
\email{mluzum@phys.washington.edu}
\affiliation{Department of Physics, University of Washington, Seattle, WA 98195-1560}
\author{Paul Romatschke}
\email{paulrom@phys.washington.edu}
\affiliation{Institute for Nuclear Theory, University of Washington, Seattle, WA 98195-1560}
\date{\today}
\preprint{INT-PUB-09-007, NT@UW-09-04}
\begin{abstract}
Hydrodynamic simulations are used to make predictions 
for the integrated elliptic flow coefficient $v_2$ in $\sqrt{s}=5.5$ 
TeV lead-lead and $\sqrt{s}=14$ TeV proton-proton collisions at the LHC.  
We predict a $10$\% increase in $v_2$ from RHIC to Pb+Pb at LHC,
and $v_2\sim 0$ in p+p collisions unless $\eta/s<0.08$.
\end{abstract}
\maketitle
\section{Introduction}
Much work has been done recently using viscous hydrodynamics to study 
the properties of gold-gold collisions at the Relativistic Heavy Ion Collider 
(RHIC)~ \cite{Luzum:2008cw,Dusling:2007gi,Song:2008si,Huovinen:2008te}.  A 
measurement of particular interest is the elliptic flow coefficient $v_2$, 
the second moment in the azimuthal angle of the distribution of emitted 
particles (cf.~\cite{Kolb:1999it}), which allows to extract information about
material constants (such as viscosity) of the high density nuclear matter created at RHIC.
Using the knowledge gained at RHIC, it should be possible to
predict experimental results at the Large Hadron Collider (LHC),
which will collide lead ions at a maximum center of 
mass energy of $\sqrt{s}=5.5$ TeV per nucleon pair compared to $\sqrt{s}=200$ GeV gold ions at RHIC.
%
If experimental data on, e.g., $v_2$ from LHC is close to 
the hydrodynamic model prediction, this would confirm that 
real progress has been made in understanding nuclear matter at extreme energy densities;
if far away, it may indicate that the successful hydrodynamic
description of experimental data from RHIC was a coincidence.

Regardless of the outcome, the advent of the RHIC experiments clearly has
lead to major progress in the theory and application of hydrodynamics to heavy-ion collisions.
A few years ago the form of the hydrodynamic equations in the presence 
of shear viscosity $\eta$ was still unresolved, with different groups 
keeping some terms while neglecting others 
\cite{Muronga:2003ta,Heinz:2005bw,Baier:2006um,Koide:2006ef}.
For the case of approximately conformal theories, where the viscosity coefficient 
for bulk---but not shear---becomes negligible, all possible terms to 
second order in gradients were derived in Ref.~\cite{Baier:2007ix}, and their relative
importance investigated in Ref.~\cite{Luzum:2008cw}. Three of the groups
performing viscous hydrodynamic simulations now agree on these terms 
\cite{Luzum:2008cw,Song:2008si,Huovinen:2008te}, 
while another group \cite{Dusling:2007gi} uses a different formalism that 
gives matching results. While this development still leaves out the consistent
treatment of bulk viscosity, the quantitative suppression of elliptic
flow by shear viscosity is therefore essentially understood. 
From comparison of viscous hydrodynamic simulations to experimental 
data \cite{Alver:2007qw,abelev:2008ed}, one can
infer an upper limit of the ratio of shear viscosity over entropy density,
$\eta/s<0.5$, for the matter produced in Au+Au collisions at 
$\sqrt{s}=200$ GeV \cite{Luzum:2008cw}, comprising extractions
by other methods \cite{Gavin:2006xd,Adare:2006nq,Drescher:2007cd}.
A sizeable uncertainty for this limit comes from the 
fact that the initial conditions for the hydrodynamic evolution are 
poorly known, with the two main models, the Glauber and Color-Glass-Condensate (CGC)
models, giving different results for the elliptic flow coefficient
\cite{Luzum:2008cw}. This difference can be understood to originate from
the different initial spatial eccentricity $e_x$ in the Glauber/CGC models. 
The eccentricity is defined as
\begin{equation}
e_x\equiv 
\frac{\left<y^2-x^2\right>}
{\left<y^2+x^2\right>}\, ,
\label{exdef}
\end{equation}
where the symbols $\left<\right>$ denote
averaging over the initial energy density in the transverse plane, $\epsilon(x,y)$.

Indeed, it had been suggested  \cite{Alt:2003ab} that the elliptic flow coefficient $v_2$
at the end of the hydrodynamic evolution would be strictly proportional to the initial 
spatial eccentricity, $v_2/e_x\propto {\rm const.}$, if the fluid was evolving
without any viscous stresses for an infinitely long time. 
This is to be contrasted with experimental data indicating
a proportionality factor of total multiplicity over
overlap area 
$v_2/e_x\propto dN/dY/S_{\rm overlap}$
\cite{Alt:2003ab}.
Total multiplicity $\frac{dN}{dY}$ here refers to the total number of 
observed particles $N$ per unit rapidity $Y$, 
while the overlap area is calculated as
\begin{equation}
S_{\rm overlap}=\pi \sqrt{\left<x^2 \right>\left<y^2\right>}\, .
\label{Sodef}
\end{equation}

Ideal fluid dynamics does not adequately describe the latest stage of 
a heavy-ion collision (the hadron gas), because of the large viscosity 
coefficient in this stage \cite{Prakash:1993bt}.
Therefore, the hydrodynamic stage lasts only for a finite time
(e.g. until all fluid cells have cooled below the decoupling temperature),
resulting in a dependence of $v_2/e_x$ on $dN/dY$. Also, viscous effects affect
the proportionality between $v_2$ and $e_x$,
leading to a behavior that is qualitatively similar to that observed in the data
\cite{Song:2008si}.

One of the objectives of this work is to extend the energy range for 
fluid dynamic results of $v_2/e_x$ from Au+Au collisions at
top RHIC to Pb+Pb collisions at top LHC energies, as well
as to study the dependence on shear viscosity. If in the future
either $e_x$ or the mean $\eta/s$ becomes known, these results can thus be used
to constrain the respective other quantity from experimental data. 
On the other hand, the values of shear viscosity 
for which the Glauber/CGC models match to experimental data at top RHIC energies 
have been extracted in Ref.~\cite{Luzum:2008cw,inprep} for Au+Au 
collisions.
Since $\eta/s$ averaged over the system evolution 
is not expected to be dramatically different for Pb+Pb collisions at the LHC,
another objective of this work is to obtain a prediction for the elliptic flow
coefficient for the LHC based on the best-fit values to RHIC.

Finally, the feasibility of detecting elliptic flow in 
p+p collisions at $\sqrt{s}=14$ TeV at the LHC is being discussed~\cite{Lpriv}.
As a reference for other approaches and experiment,
it interesting to study the possible size and viscosity dependence of $v_2$ 
under the hypothetical assumption that the bulk evolution following p+p collisions could be 
captured by fluid dynamics.

\section{Setup}
To make predictions for nuclear collisions at LHC energies, we use our hydrodynamic
model that successfully described experimental data at RHIC \cite{Luzum:2008cw,inprep}
and make modifications 
to the input parameters appropriate for the higher collision energies at the LHC.

As a reminder, the hydrodynamic model \cite{Luzum:2008cw} is based on the 
conservation of the energy momentum tensor
\cite{Baier:2007ix}
\begin{eqnarray}
T^{\mu \nu}&=&\epsilon u^\mu u^\nu- p \Delta^{\mu \nu}+\Pi^{\mu \nu}\, ,\nonumber\\
\Pi^{\mu\nu} &=& \eta \nabla^{\langle \mu} u^{\nu\rangle}
- \tau_\Pi \left[ \Delta^\mu_\alpha \Delta^\nu_\beta D\Pi^{\alpha\beta} 
 + \frac 4{3} \Pi^{\mu\nu}
    (\nabla_\alpha u^\alpha) \right] \nonumber\\
  &&\hspace*{-1cm} -\frac{\lambda_1}{2\eta^2} {\Pi^{<\mu}}_\lambda \Pi^{\nu>\lambda}
  +\frac{\lambda_2}{2\eta} {\Pi^{<\mu}}_\lambda \omega^{\nu>\lambda}
  - \frac{\lambda_3}{2} {\omega^{<\mu}}_\lambda \omega^{\nu>\lambda}\, ,
\nonumber
\end{eqnarray}
where $\epsilon,p$ and $u^\mu$ are the energy density, pressure,
and fluid 4-velocity, respectively. $D\equiv u^\mu D_\mu$ and 
$\nabla_\alpha\equiv \Delta_\alpha^\mu D_\mu$
are time-like and space-like projections of the covariant derivative $D_\mu$,
where $\Delta^{\mu \nu}=g^{\mu \nu}-u^\mu u^\nu$
and we remind the compact notations 
$A_{\langle \mu} B_{\nu\rangle} \equiv \left(\Delta^\alpha_\mu \Delta^\beta_\nu + 
\Delta^\alpha_\nu \Delta^\beta_\mu-\frac{2}{3} \Delta^{\alpha \beta} 
\Delta_{\mu \nu}\right) A_\alpha B_\beta$ and 
$\omega_{\mu \nu}\equiv \frac 1 2 \left( \nabla_\nu u_\mu - \nabla_\mu u_\nu \right)$.
For relativistic nuclear
collisions it is convenient to follow Bjorken \cite{Bjorken:1982qr} 
and use Milne coordinates proper time 
$\tau=\sqrt{t^2-z^2}$ and spacetime rapidity $\xi={\rm atanh}\frac{z}{t}$,
in which the metric becomes $g_{\mu \nu}={\rm diag}(1,-1,-1,-\tau^2)$,
and assume that close to 
$\xi=0$, the hydrodynamic
degrees of freedom are approximately boost-invariant ($\xi\simeq Y$).

The hydrodynamic equations $D_\mu T^{\mu \nu}=0$ 
then constitute an initial value problem in proper time 
and transverse space, and are solved numerically 
(see Ref.~\cite{Luzum:2008cw}). The input parameters for 
hydrodynamic evolution are the equation of state $p=p(\epsilon)$ 
and the first (second) order hydrodynamic transport coefficients $\eta$
($\tau_\Pi,\lambda_1,\lambda_2,\lambda_3$).
The values for $\lambda_{1,2,3}$ have been found 
to hardly affect the boost-invariant hydrodynamic evolution for Au+Au collisions
at RHIC \cite{Luzum:2008cw}, so here they are generally set to zero.

\begin{table}
\begin{tabular}{|cccccc|}
\hline\hline
Beam &
Initial cond. & 
$\frac{dN_{\rm ch}}{dY}$ & 
$T_i$ [GeV] & 
$\sqrt{s}$ [GeV] &
$\tau_0$ [fm/c] \\
\hline
Gold& Glauber& 800 & 0.34 & 200&1\\
Gold& CGC& 800 & 0.31 & 200&1\\
Lead& Glauber& 1800& 0.42& 5500&1\\
Lead& CGC& 1800& 0.39& 5500&1\\
Protons& Glauber& 6& 0.400& 14000& 0.5\\
Protons& Glauber& 6& 0.305& 14000& 1\\
Protons& Glauber& 6& 0.270& 14000& 2\\
\hline\hline
\end{tabular}
\caption{Central collision parameters used for the viscous hydrodynamics
simulations ($T_f=0.14$ GeV for all).
}
\label{tab:par}
\end{table}

The mechanisms leading to thermalization (the onset of hydrodynamic behavior) 
are not well understood in nuclear collisions.  Therefore, 
it is not known 
how the thermalization time $\tau_0$ at which hydrodynamic evolution 
is started depends on the collision energy.  Barring further insight, 
we start hydrodynamic evolution for the LHC at the same
time as for the RHIC simulations ($\tau_0=1$ fm/c). At this time,
%
the initial conditions for the transverse
energy density $\epsilon(x,y)$ are given by the Glauber or CGC model, respectively,
the fluid velocities are assumed to vanish, and the shear tensor $\Pi^{\mu \nu}$
is set to zero (other values for $\Pi^{\mu \nu}$ do not seem to
affect the final results \cite{Song:2007ux,Luzum:2008cw}). 
For brevity, we refer to Ref.~\cite{Luzum:2008cw} for the details of
the Glauber and CGC models, but for the expert reader note
that we use the Woods-Saxon parameters of radius $R_0=6.4\, (6.6)$ fm and skin 
depth $\chi=0.54\,(0.55)$ fm  for gold (lead), 
and assume a nucleon-nucleon cross section of $\sigma= 40\,(60)$ mb 
for $\sqrt{s}=200\,(5500)$ GeV collisions.

The overall normalization
of the initial energy density (parametrized by the initial temperature at the center $T_i$) 
was adjusted to match the experimentally observed
multiplicity at RHIC; by analogy, for LHC the normalization
is adjusted to match the predicted multiplicity 
\cite{Kharzeev:2004if,Armesto:2008fj,Abreu:2007kv,Busza:2007ke}. 
Since we lack detailed knowledge about its temperature dependence, 
the ratio of shear viscosity to entropy density $\eta/s$ is set to be constant
during the hydrodynamic evolution
(equal to the average over the spacetime evolution of the system).
The relaxation time coefficient $\tau_\Pi$ is expected \cite{Baier:2007ix,York:2008rr}
to lie in the range
$\frac{\tau_\Pi}{\eta} (\epsilon+p)\simeq2.6-6$.
The equation of state (EoS) can in principle be provided by lattice QCD.
While at present there are points of disagreement between lattice groups about, e.g.,  
the precise location of the QCD phase transition, 
there is consensus that it is an analytic crossover 
\cite{Aoki:2009sc, Bazavov:2009zn}.  Therefore, we use a
lattice-inspired EoS \cite{Laine:2006cp} 
that is consistent with both the current 
consensus and perturbative QCD; also, since it resembles \cite{Bazavov:2009zn},
we expect that using a different lattice EoS will have a 
minor effect on our results.

Once a given fluid cell has cooled down to the decoupling temperature
$T_f$, its energy and momentum are converted into particle
degrees of freedom using the Cooper-Frye freeze-out prescription \cite{Cooper:1974mv}.
A value of $T_f=0.14$ GeV was determined by matching to RHIC data 
and will also be used for LHC energies,
assuming that it is mostly determined by local conditions, and
less so by initial energy density, system size or collision energy.
\begin{figure}
\includegraphics[width=\linewidth]{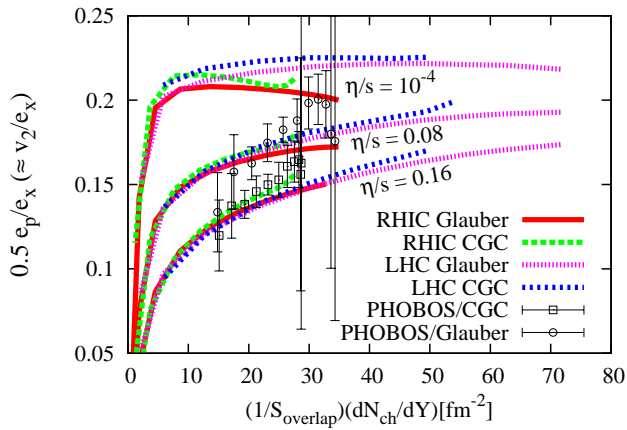}
\caption{\label{plot1}
(Color online) Anisotropy (\ref{v2def}) divided by (\ref{exdef}), as a function of 
initial entropy (\ref{svsn}) divided by (\ref{Sodef}).
Shown are results from hydrodynamic simulations for $\sqrt{s}=200$ GeV Au+Au 
(RHIC) and $\sqrt{s}=5.5$ TeV Pb+Pb collisions (LHC). 
For comparison, experimental data for $v_2$ from RHIC \cite{Alver:2006wh}, 
divided by
$e_x$ from two models \cite{Drescher:2007cd},
is shown as a function of measured
$\frac{dN_{\rm ch}} {dY}$ \cite{Adler:2003cb} divided by (\ref{Sodef}). See text for details.}
\end{figure}
The distribution of the particle degrees of freedom may be further
evolved using a hadronic cascade code (as in Ref.~\cite{Bass:2000ib}), or
in a more simple approach the unstable particle resonances
are allowed to decay, without further evolving the stable particle
distributions. In both cases, the total multiplicity and particle
correlations (such as the elliptic flow coefficient) are then calculated
from the stable particle distribution (cf.~\cite{Luzum:2008cw}). 
Surprisingly, it was found in Ref.~\cite{Kolb:1999it,Luzum:2008cw} that 
the momentum integrated elliptic flow coefficient for charged hadrons---to good approximation---is equal to half the momentum anisotropy,
\begin{equation}
v_2\simeq \frac{1}{2} e_p=\frac{1}{2}\frac{\int dx dy\, T^{xx}-T^{yy}}{\int dx dy\, T^{xx}+T^{yy}}\, .
\label{v2def}
\end{equation}
Since
the momentum anisotropy is a property of the fluid, it is independent on
the details of the freeze-out procedure and only mildly dependent on the choices of
$\tau_0,T_f$.
Unlike at RHIC where pairs of $\tau_0$ and $T_f$ could be fine-tuned to fit the particle spectra
at central collisions, no such extra information is available for the LHC.
Hence Eq.~(\ref{v2def}) may provide the most reliable way of determining
the elliptic flow of charged hadrons, and will be used in the following.
Similarly, one can use the total entropy per unit spacetime rapidity $\frac{dS}{d\xi}$
in the fluid as a proxy for the total (charged hadron) multiplicity per unit rapidity 
$\frac{dN}{dY}$
($\frac{dN_{\rm ch}}{dY}$)
with a proportionality factor \cite{Gyulassy:1983ub,Kolb:2001qz}
\begin{equation}
\frac{d S}{d\xi}\sim\frac{d S}{dY}\simeq 4.87 \frac{dN}{dY}\simeq 7.85 \frac{dN_{\rm ch}}{d Y}.
\label{svsn}
\end{equation}
Note that for a gas of massive hadrons in thermal equilibrium 
at $T_f=0.14$ GeV the ratio of entropy to particle density is 
$\sim 6.41$, but the decay of unstable resonances 
produces additional entropy, resulting in Eq.~(\ref{svsn}).
Since results from RHIC suggest there is only approximately $10$\% viscous entropy production
during the hydrodynamic phase \cite{Romatschke:2007jx,Song:2008si}, the entropy 
$\frac{d S}{dY}$ at $\tau=\tau_0$ can be used
to estimate the final particle multiplicity. In the case of the LHC,
the world average for the predicted charged hadron multiplicity 
for central Pb+Pb collisions at $\sqrt{s}=5.5$ TeV \cite{Armesto:2008fj},
$\frac{dN_{\rm ch}}{d Y}\simeq 1800$, can be used to estimate the 
total entropy at $\tau=\tau_0$, and hence the overall normalization $T_i$
of the initial energy density (see Tab.~\ref{tab:par}). 

Using Eqs.~(\ref{v2def},\ref{svsn}) for the multiplicity and elliptic flow
allows to make predictions for the LHC without having to model
the hadronic freeze-out, which should make the results more robust.
However, as a consequence one does not get information about
the momentum dependence of the elliptic flow coefficient, prohibiting
detailed comparison with predictions by other groups 
\cite{Chaudhuri:2008je,Abreu:2007kv}.

\begin{figure}
\includegraphics[width=\linewidth]{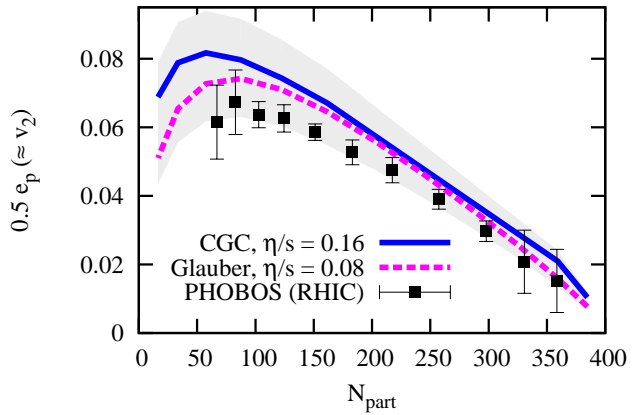}
\caption{\label{plot2}
(Color online) Anisotropy (\ref{v2def}) prediction for $\sqrt{s}=5.5$ TeV Pb+Pb collisions (LHC),
as a function of centrality.
Prediction is based on values of $\eta/s$ for the Glauber/CGC model that
matched $\sqrt{s}=200$ GeV Au+Au collision data from PHOBOS
at RHIC (\cite{Alver:2006wh},  shown for comparison).  The shaded band corresponds to the estimated uncertainty in our prediction from additional systematic effects: using $e_p/2$ rather than $v_2$ (5\%) \cite{Luzum:2008cw}; using a lattice EoS from \cite{Laine:2006cp} rather than \cite{Bazavov:2009zn} (5\%); not including hadronic cascade afterburner (5\%) \cite{Teaney:2001av}}.
\end{figure}

\section{Results}

With the initial energy density distribution fixed at $\tau_0$,
the hydrodynamic model then gives predictions for the ratio
of $v_2/e_x$ at the LHC. In Fig.~\ref{plot1}, the results are shown
for three different values of shear viscosity, for two different
initial conditions and two different beams/collision 
energies (Au+Au at $\sqrt{s}=200$ GeV, Pb+Pb at $\sqrt{s}=5.5$ TeV). 
The resulting values for
$v_2/e_x$ seem to be quasi-universal functions of the total multiplicity scaled
by the overlap area $S_{\rm overlap}$, only depending on the value of $\eta/s$
(and, to a lesser extent, the collision energy). The deviations of the RHIC
simulations from the universal curve can be argued to arise from 
a combination of the finite lifetime of the hydrodynamic phase at $\sqrt{s}=200$ GeV 
and the presence of the QCD phase transition, and is strongest 
for ideal hydrodynamics, in agreement with earlier findings \cite{Song:2008si}.

Also shown in Fig.~\ref{plot1} is experimental data for the elliptic flow coefficient
for Au+Au collisions at RHIC, normalized
by $e_x$ from a Monte-Carlo calculation (including fluctuations) 
in Glauber and CGC models 
(see Ref.~\cite{Drescher:2007cd} for details). Since $e_x$ is not 
directly measurable, the differently normalized data gives an estimate
of the overall size of $v_2/e_x$ at RHIC. Directly matching experimental
data on $v_2$ using a hydrodynamic model with an initial $e_x$
specified by the Glauber or CGC model, a reasonable fit was achieved
for a mean value of $\eta/s\simeq0.08$ and $\eta/s\simeq0.16$, respectively 
\cite{Luzum:2008cw,inprep}. Under the assumption that the average $\eta/s$ is similar for
collisions at RHIC and the LHC (along with the assumptions 
discussed above), one can make a prediction for the integrated elliptic flow
coefficient for charged hadrons as a function of impact parameter 
(or more customarily the number of participants $N_{\rm part}$, cf.~\cite{Luzum:2008cw}).
The result is shown in  Fig.~\ref{plot2}. As can be seen, we expect
integrated $v_2$ at the LHC to be about ten percent larger than at RHIC, which is less
than the prediction by ideal hydrodynamics \cite{Niemi:2008ta},
and in agreement with the extrapolations by Drescher et al.~\cite{Abreu:2007kv}.

Finally, using the charge density parametrization of the proton $\rho(b)$ 
in Ref.~\cite{Miller:2007uy} as an equivalent of 
the nuclear thickness function in the Glauber model (cf.~\cite{Luzum:2008cw})
one obtains an estimate for the shape of the transverse energy density
following a relativistic p+p collision. Using the 
predicted multiplicity at mid-rapidity $\frac{dN}{dY}\sim 6$ 
\cite{Kharzeev:2004if,Busza:2007ke} for $\sqrt{s}=14$ TeV p+p collisions at the LHC, one 
can again use Eq.~(\ref{svsn}) to infer the overall normalization
of the energy density (or $T_i$) at $\tau=\tau_0$ (see Tab.~\ref{tab:par}).
As a ``Gedankenexperiment'' one can then ask how much elliptic flow
would be generated in LHC p+p collisions if the subsequent evolution
was well approximated by boost-invariant viscous hydrodynamics.
One finds that for ideal hydrodynamics $\frac{e_p}{2}\sim v_2\sim 0.035$ 
for integrated $|v_2|$ in minimum bias collisions (cf.~(28) in \cite{Luzum:2008cw}), 
while for $\eta/s\ge 0.08$, $v_2$ typically changes by 
almost 100 percent when varying the relaxation time
$\frac{\tau_\Pi}{\eta} (\epsilon+p)$ between $2.6$ and $6$ and varying
$\tau_0$ by a factor of two.
This indicates that for $\eta/s\ge 0.08$, the hydrodynamic
gradient expansion does not converge and as a consequence 
it is unlikely that 
elliptic flow develops in p+p collisions at top LHC energies.
If experiments find a non-vanishing value for integrated $|v_2|>0.02$ in 
minimum bias p+p collisions, 
this would be an indication for an extremely small viscosity $\eta/s<0.08$
in deconfined nuclear matter.

To conclude, viscous hydrodynamics can be used to make predictions
for the ratio of $v_2/e_x$ as a function of multiplicity and $\eta/s$. Assuming
a multiplicity of $\frac{dN_{\rm ch}}{d Y}\simeq 1800$ 
for the matter created in Pb+Pb collisions at LHC, as well as $\eta/s$
similar to RHIC, we predict the integrated
elliptic flow for charged hadrons to be $10$\% larger at the LHC than at RHIC.
We expect $v_2$ measurements in p+p collisions to be consistent with zero,
unless the shear viscosity is extremely small ($\eta/s<0.08$).

\begin{acknowledgments}
We would like to thank J.-Y.~Ollitrault for providing tabulated results from
\cite{Drescher:2007cd}, and J.~Albacete, K.~Eskola, G.A.~Miller, L.~Ramello,
J.~Schukraft, R.~Snellings and P.~Steinberg for discussions.
The work of ML and PR was supported by the US Department of Energy, grant 
numbers DE-FG02-97ER41014 and DE-FG02-00ER41132, respectively.
\end{acknowledgments}

\bibliographystyle{apsrev}
\bibliography{LHCpaper}

\end{document}